# Rotational spin Hall effect in a uniaxial crystal

T. Fadeyeva, C. Alexeyev, A. Rubass, M. Ivanov, A. Zinov'ev,

V. Konovalenko and A. Volyar[*]

*Department of Physics, Taurida National V.I. Vernadsky University*

*4, Vernadsky av. Simferopol, Ukraine, 95007*

[*]*Corresponding author:* <volyar@crimea.edu>

***Abstract*** *We have considered the propagation process of the phase-matched array of singular beams through a uniaxial crystal. We have revealed that local beams in the array are rotated when propagating. However the right and left rotations are unequal. There are at least two processes responsible for the array rotation: the interference of local beams and the spatial depolarization. The interference takes place in the vortex birth and annihilation events forming the symmetrical part of the rotation. The depolarization process contributes to the asymmetry of the rotation that is called the rotational spin Hall effect. It can be brought to light due to the difference between the envelopes of the dependences of the angular displacement on the inclination angle of the local beams or the crystal length reaching the value some angular degree. The direction of the additional array rotation is exclusively defined by the handedness of the circular polarization in the initial beam array.*

*OCIS codes:* 350.5030, 260.6042, 260.1180, 260.0260

## I. Introduction

Transmission and reflection of bounded light beams at the boundary face of two dielectric media is accompanied by deflection of the beam axis from the geometrical direction, namely, the beam axis is shifted laterally relative to the initial plane of reflection. For the first time this phenomenon was predicted by Fedorov [1] and experimentally proved by Imbert [2]. In essence, the effect manifests itself in vicinity of a thin boundary layer and therefore it is of the local action. The effect also is inherent in all paraxial beams (see, e.g. [3]) being controlled by the energy flux circulation in vicinity of the boundary face [4, 5] and the conservation law of the total angular momentum flux [6]. Later on similar optical phenomena accompanied by the lateral shift of the beams came to be called the spin Hall effect [7-12]. Apart from the lateral shift there takes place also a weak angular displacement in the reflected and transmitted light [13, 14], the shift value being also defined by the sign and value of the topological charge of the optical vortex carried by the beam. On the other hand, non-homogeneous media opened a new side of the spin Hall effect – its nonlocallity. The complex light beam propagating along a non-homogeneous medium rotates around the axis of symmetry of the medium (e.g. of the optical fiber) [12, 15, 16]. But the most interesting is that the beam rotation depending on the handedness of the circular polarization state is observed even in free space [17, 18] for nonparaxial beams.

Recently we have revealed that a homogeneous birefringent medium of a uniaxial crystal causes the nonlocal spin Hall effect [19-21]. The circularly polarized paraxial beam tilted relative to the crystal optical axis proves to be laterally shifted by the distance proportional to the wavelength and inversely proportional to the inclination angle of the beam axis. The basic process that controls the beam propagation and, in particular, the lateral shift in the crystal is the spin-orbit coupling that regulates energy exchange between the spin and orbital angular momentum fluxes. The beam is spatially depolarized at a large crystal length, its spin angular momentum vanishes whereas the total angular momentum flux as a sum of the spin and orbit angular momenta must be conserved. As a result, the asymptotic value of the lateral shift depends neither on the type of the paraxial vortex-beam nor on its topological charge defined only by the handedness of the circular polarization state of the initial beam at the crystal input.

However, the array of tilted paraxial beams in a uniaxial crystal shaped in the form of the axially symmetric set of the light beams whose phases are matched can break such a simple developing of the wave process down. Indeed, the symmetric array of singular beams in free space forms at the beam axis the optical vortex whose topological charge is defined by the topological charges $l$ of each partial beam, the phase-matching number $m$ of the beam array and the number $N$ of partial (or local) beams [22, 23]. Besides, when increasing the beam number in the array $(N \to \infty)$ the beam behavior comes to be similar to that of ordinary co-axial paraxial beam [24] (in particular, if the topological charge $l = 0$ of the beam array is transformed into the Bessel-Gaussian beam [25, 26]). It means that the circularly polarized beam array propagation will result in generation of the doubly charged optical vortex at the array axis [20]. Any lateral shifts of the partial beams have to vanish lest the conservation law of the angular momentum flux is broken down. However, at the intermediate range of small inclination angles and a limited number of partial beams the interferential processes between the partial beams have to turn the lateral shift of each partial beam into a rotational movement of the beam array as a whole. This process we will call later on the nonlocal rotational spin Hall effect .

The aim of our paper is to study theoretically and experimentally the competition processes that form and break down the rotational spin Hall effect.

## II. Phase-matched array of singular beams in a uniaxial crystal

### II.1 The basic equations and major properties

Let us consider at first the properties of the beam array consisting of the $N$ paraxial local beams in a birefringent crystal with the permittivity tensor $\hat{\varepsilon} = diag(\varepsilon_o, \varepsilon_o, \varepsilon_e)$, $\varepsilon_o = n_o^2$, $\varepsilon_e = n_3^2$, $n_e = n_3^2 / n_o$, $n_{o,e}$ are ordinary and extraordinary refractive indices [19,20]. The axes of local beams in the array are positioned at the vertices of a regular polygon (see Fig.1). The axis of the ordinary beam is shifted by the distance $r_0$ relative to the origin of the referent frame of the array and is tilted at the angle $\alpha_o$ to the z-axis of the array (by the angle $\alpha_e$ for the extraordinary beam). Before describing the beam array in the global coordinates $\{x, y, z\}$, one writes the electric field of the scalar ordinary vortex-beam $\Psi_o$ with the topological charge $l$ in the local coordinates $\{x'_n, y'_n, z'_n\}$ with $\alpha_o = 0$:

$$\Psi_o = \left[\frac{x'_n - i y'_n}{w_0 \sigma_o}\right]^l \exp\left(-\frac{x'^2_n + y'^2_n}{w_0^2 \sigma_o}\right) / \sigma_o, \tag{1}$$

$\sigma_o = 1 - iz/z_o$, $z_o = k_o w_0^2 / 2$, $w_0$ is the beam waist radius at the $z = 0$. The inclination angles $\alpha_{o,e}$ can be taken into account by a simple substitution [20] $y'_n \rightarrow y'_n + i\alpha_o z_o$. It should be noted that there are two mode beam groups in a uniaxial crystal [27].

The right and left hand polarized components (RHP and LHP) of the first group in the local coordinates can be written now in the form

$$\left(E_+^{-l}\right)_{loc} = \left[\frac{x'_n - i(y'_n - \alpha_o z)}{w_0 \sigma_o}\right]^l \tilde{\Psi}_o + \left[\frac{x'_n - i(y'_n - \alpha_e z)}{w_0 \sigma_e}\right]^l \tilde{\Psi}_e, \tag{2}$$

$$\left(E_-^{-l}\right)_{loc} = -\sum_{p=0}^{l} \binom{l}{p}\left(-\frac{\alpha_o z_o}{w_0}\right)^{l-p} \left[\left(\frac{x'_n - i(y'_n + i\alpha_o z_o)}{w_0 \sigma_o}\right)^{l-2}\left(\frac{l-1}{\sigma_o} - \frac{r_n^2}{w_0^2 \sigma_o^2}\right)\tilde{\Psi}_o - \right.$$
$$\left. -\left(\frac{x'_n - i(y'_n + i\alpha_o z_o)}{w_0 \sigma_e}\right)^{l-2}\left(\frac{l-1}{\sigma_e} - \frac{r_n^2}{w_0^2 \sigma_e^2}\right)\tilde{\Psi}_e\right], \tag{3}$$

where $\sigma_e = 1 - iz/z_e$, $z_e = k_e w_0^2 / 2$, $k_{o,e} = n_{o,e} k_0$, $k_0$ is a wavenumber in free space,

$$\tilde{\Psi}_{o,e} = \exp\left(-\frac{x^2 + (y + i\alpha_o z_o)^2}{w_0^2 \sigma_{o,e}} - k_o \frac{\alpha_o^2 z_o}{2}\right) / \sigma_{o,e}. \tag{4}$$

The electric field of the second group of the local beams can be written as

$$\left(E_+^{l}\right)_{loc} = \left[\frac{x'_n + i(y'_n - \alpha_o z)}{w_0 \sigma_o}\right]^l \tilde{\Psi}_o + \left[\frac{x'_n + i(y'_n - \alpha_e z)}{w_0 \sigma_e}\right]^l \tilde{\Psi}_e, \tag{5}$$

$$\left(E_{-}^{l}\right)_{loc} = \sum_{p=0}^{l}\binom{l}{p}\left(\frac{\alpha_o z_o}{w_0}\right)^{l-p}\left(\frac{x_n' + i\left(y_n' + i\alpha_o z_o\right)}{w_0}\right)^{p+2}\left[\sum_{j=0}^{p+1}\frac{(p+1)!}{j!(\sigma_o)^j}\left(\frac{r_n}{w_0}\right)^{2(j-p-2)}\tilde{\Psi}_o - \right.$$
$$\left. - \sum_{j=0}^{p+1}\frac{(p+1)!}{j!(\sigma_e)^j}\left(\frac{r_n}{w_0}\right)^{2(j-p-2)}\tilde{\Psi}_e\right], \tag{6}$$

where $r_n^2 = x_n'^2 + \left(y_n' + i\alpha_o z_o\right)^2$.

The transition to the global coordinates is defined by three points: the first, the coordinate transformation:

$$x_n' = x\cos\varphi_n + y\sin\varphi_n + r_0, \quad y_n' = -x\sin\varphi_n + y\cos\varphi_n; \tag{7}$$

the second, the transformation of the circularly polarized basis:

$$\hat{\mathbf{e}}_\pm = \left(\hat{\mathbf{x}}_n \pm i\hat{\mathbf{y}}_n\right)\exp\left(\mp i\varphi_n\right), \quad \varphi_n = \frac{2\pi}{N}n \tag{8}$$

and, the third, the summation of the local beams. Besides, the phase-matched beam array is formed in such a way that each local beam has its own phase $\phi_{n,m} = \frac{2\pi n}{N}m$, where $m$ is a real number. However, the local beams in the obtained construction have additional phases $\exp(i\varphi_n)$ at the initial z=0 plane, caused by the basis transformation (8). But our requirement is that the RHP component $E_+(z=0)$ of the local beams in the array is phase- matched at the z=0 plane by the relation $\phi_{n,m} = \frac{2\pi n}{N}m$ (while $E_-(z=0)=0$). Thus we have to multiply the above equations by the factor $\exp(i\varphi_n)$. Finally we obtain the expressions for the first mode group:

$$E_+^{-l,m} = E_o^{l,m} + E_e^{l,m} = \sum_{n=1}^{N}\left\{\left[\frac{x_n' - i\left(y_n' - \alpha_o z\right)}{w_0\sigma_o}\right]^l\tilde{\Psi}_o + \left[\frac{x_n' - i\left(y_n' - \alpha_e z\right)}{w_0\sigma_e}\right]^l\tilde{\Psi}_e\right\}e^{im\varphi_n}, \tag{9}$$

$$E_-^{-l,m} = -\sum_{n=1}^{N}\left\{\sum_{p=0}^{l}\binom{l}{p}\left(-\frac{\alpha_o z_o}{w_0}\right)^{l-p}\left[\left(\frac{x_n' - i\left(y_n' + i\alpha_o z_o\right)}{w_0\sigma_o}\right)^{l-2}\left(\frac{l-1}{\sigma_o} - \frac{r_n^2}{w_0^2\sigma_o^2}\right)\tilde{\Psi}_o - \right.\right.$$
$$\left.\left. - \left(\frac{x_n' - i\left(y_n' + i\alpha_o z_o\right)}{w_0\sigma_e}\right)^{l-2}\left(\frac{l-1}{\sigma_e} - \frac{r_n^2}{w_0^2\sigma_e^2}\right)\tilde{\Psi}_e\right]\right\}e^{i(2+m)\varphi_n}, \tag{10}$$

The orthogonal electric field for the first group has the form

$$E_+^{-l,m} = \sum_{n=1}^{N}\left\{\sum_{p=0}^{l}\binom{l}{p}\left(-\frac{\alpha_o z_o}{w_0}\right)^{l-p}\left[\left(\frac{x'_n+i(y'_n+i\alpha_o z_o)}{w_0\sigma_o}\right)^{l-2}\left(\frac{l-1}{\sigma_o}-\frac{r_n^2}{w_0^2\sigma_o^2}\right)\tilde{\Psi}_o\right.\right.$$

$$\left.\left.-\left(\frac{x'_n+i(y'_n+i\alpha_o z_o)}{w_0\sigma_e}\right)^{l-2}\left(\frac{l-1}{\sigma_e}-\frac{r_n^2}{w_0^2\sigma_e^2}\right)\tilde{\Psi}_e\right]\right\}e^{i(-2+m)\varphi_n}, \quad (11)$$

$$E_-^{-l,m} = \sum_{n=1}^{N}\left\{\left[\frac{x'_n+i(y'_n-\alpha_o z)}{w_0\sigma_o}\right]^l\tilde{\Psi}_o+\left[\frac{x'_n+i(y'_n-\alpha_e z)}{w_0\sigma_e}\right]^l\tilde{\Psi}_e\right\}e^{im\varphi_n}. \quad (12)$$

In the above equations (11) and (12) we have multiplied the electric field by the factor $\exp(-i\varphi_n)$ because $E_+(z=0)=0$.

The phase-matched array of the second mode group is written as

$$E_+^{l,m} = \sum_{n=1}^{N}\left\{\left[\frac{x'_n+i(y'_n-\alpha_o z)}{w_0\sigma_o}\right]^l\tilde{\Psi}_o+\left[\frac{x'_n+i(y'_n-\alpha_e z)}{w_0\sigma_e}\right]^l\tilde{\Psi}_e\right\}e^{im\varphi_n}, \quad (13)$$

$$E_-^{l,m} = \sum_{n=1}^{N}\left\{\sum_{p=0}^{l}\binom{l}{p}\left(\frac{\alpha_o z_o}{w_0}\right)^{l-p}\left(\frac{x'_n+i(y'_n+i\alpha_o z_o)}{w_0}\right)^{p+2}\left[\sum_{j=0}^{p+1}\frac{(p+1)!}{j!(\sigma_o)^j}\left(\frac{r_n}{w_0}\right)^{2(j-p-2)}\tilde{\Psi}_o\right.\right.$$

$$\left.\left.-\sum_{j=0}^{p+1}\frac{(p+1)!}{j!(\sigma_e)^j}\left(\frac{r_n}{w_0}\right)^{2(j-p-2)}\tilde{\Psi}_e\right]\right\}e^{i(2+m)\varphi_n}. \quad (14)$$

The orthogonal electric field is

$$E_+^{l,m} = -\sum_{n=1}^{N}\left\{\sum_{p=0}^{l}\binom{l}{p}\left(\frac{\alpha_o z_o}{w_0}\right)^{l-p}\left(\frac{x'_n-i(y'_n+i\alpha_o z_o)}{w_0}\right)^{p+2}\left[\sum_{j=0}^{p+1}\frac{(p+1)!}{j!(\sigma_o)^j}\left(\frac{r_n}{w_0}\right)^{2(j-p-2)}\tilde{\Psi}_o\right.\right.$$

$$\left.\left.-\sum_{j=0}^{p+1}\frac{(p+1)!}{j!(\sigma_e)^j}\left(\frac{r_n}{w_0}\right)^{2(j-p-2)}\tilde{\Psi}_e\right]\right\}e^{i(-2+m)\varphi_n}. \quad (15)$$

$$E_-^{l,m} = \sum_{n=1}^{N}\left\{\left[\frac{x'_n-i(y'_n-\alpha_o z)}{w_0\sigma_o}\right]^l\tilde{\Psi}_o+\left[\frac{x'_n-i(y'_n-\alpha_e z)}{w_0\sigma_e}\right]^l\tilde{\Psi}_e\right\}e^{im\varphi_n}. \quad (16)$$

In the above equations we made use of the matching condition: $k_o\alpha_o=k_e\alpha_e$ for the ordinary and extraordinary local beams.

The equations obtained enable us to describe the basic features of the beam array in the crystal. In our later consideration, to fix the idea, we chose the equations (9)-(12) in such a way that they characterize the beams at the $z=0$ plane with opposite signs of the vortex topological charge and handedness of the circular polarization. These expressions permit to shape a great number of beam fields with space-variant polarization and intricate intensity distributions that can be appropriate for practical devices that deal not only with technical [26, 28, 29] but also with biological [30, 31] objects. Typical intensity distributions of the beam arrays are shown in Fig.2 where we have used designation of the array state in the form $\{N,m,l\}$ [22]. The field structure can be gradually changed by tuning the lateral displacement of the local beams $r_0$ in the range of fractions of wavelength and the polarization state of the array after the crystal.

It is significant to note that the phase-matched beam array can be transformed into the spiral beam [23, 24] under the condition $r_0 = \alpha_o z_o = \alpha_e z_e$. In the general case, the array propagation is accompanied by structural transformations of the array electric field as it is shown in Fig.3A. At the same time, the spiral beams are of the stable field structures under their propagation up to the scale and rotation (see Fig.3B). Note that the transition from the RHP state into the LHP one permits in a lot of cases to highlight the outer contours of the beam array as it is demonstrated by the lowest line of the patterns in Fig.3B that can be important for numerous practical applications.

*II.2. The topological charge of the centered optical vortex*

In a great number of practical cases the centered optical vortex in the polarized components of the cylindrically symmetric beam array contains the basic information about the orbital angular momentum (OAM) [23]. But in order to get round to this information it is necessary to transform radically the expressions (9) and (10). Because the basic mathematical procedures for the scalar beam array are described in the paper [23, 32] we remark here only some peculiarities concerning the vector approach. For simplicity we restrict ourselves to the case of the Gaussian local beams $(l=0)$. First of all, the RHP components of the ordinary and extraordinary beam arrays have the same centered optical vortices and we can use the results of the papers [23, 33] *in corpore* here. To transform the LHP component we make at first some approximations. We consider the area near the optical axis where $r^2 = x^2 + y^2 \ll r_0^2, \alpha_o^2 z_o^2$. Since

$$r_n^2 = r^2 + r_0^2 + 2r_0 r \cos(\varphi - \varphi_n) - 2i\alpha_o z_o r \sin(\varphi - \varphi_n) - \alpha_o^2 z_o^2 \tag{17}$$

we can approximately write: $1/r_n^2 \approx -1/(\alpha_o z_o)^2$ in assuming $r_0 = 0$. Besides,

$$\left[x_n - i(y_n + i\alpha_o z_o)\right]^{-2} = r_n^{-4}\left[x_n + i(y_n + i\alpha_o z_o)\right]^2 \approx \left[x_n + i(y_n + i\alpha_o z_o)\right]^2 / (\alpha_o z_o)^4.$$

Then we can use the generatrix expression for the modified Bessel functions $I_n(x)$:

$$\exp\left[\left(t + \frac{1}{t}\right)\frac{x}{2}\right] = \sum_{k=-\infty}^{\infty} t^k I_k(x), \tag{18}$$

bearing in mind that $I_k(x) \approx (x/2)^k / k!$ for $x \ll 1$, and the expression for the geometric progression:

$$\sum_{n=1}^{\infty} \exp\left\{i(m-p)\frac{2\pi}{N}n\right\} = \begin{cases} N, \text{if } m - p = qN, q = 0, \pm 1, \pm 2, \ldots \\ 0, \quad \text{otherwhere} \end{cases}. \tag{19}$$

Keeping the terms with $1/(\alpha_o z_o)^0$ in eq. (10) we come to

$$E_- \sim \sum_{q=-\infty}^{\infty} (-1)^{(qN+m)/2} \exp\{i[2 + qN + 2m]\varphi\} \left[I_{qN+2m+2}(\xi_o)\Psi_o - I_{qN+2m+2}(\xi_e)\Psi_e\right], \tag{20}$$

where $\Psi_{o,e} = \exp\left(-\frac{r^2}{w_0^2 \sigma_{o,e}}\right) / \sigma_{o,e} e^{-ik_{o,e}z} \exp\left(-\frac{k_o z_o \alpha_o^2}{2}\right) \exp\left[-\frac{\bar{\alpha}^2 - \Re^2}{2\sigma_{o,e}}\right]$, $\xi_{o,e} = 2\frac{z_{o,e}\sqrt{\bar{\alpha}^2 - \Re^2}}{z + i z_{o,e}}R$,

$\Re = r_0 / w_0$, $\bar{\alpha} = \alpha_{o,e} / \alpha_{dif}^{(o,e)}$, $\alpha_{dif}^{(o,e)} = 2/(k_{o,e}w_0)$. $R = r/w_0$.

Obviously, the major contribution to the series (20) gives the terms with $q = 0, \pm 1$ whence we obtain the simple rule for the value of the centered topological charge $Q_-$, if $\alpha_o > 0$:

$$Q_- = \begin{cases} m+2, & N > 2(m+2) \\ m+2-N, & N < 2(m+2) \\ 0, & N = 2(m+2) \end{cases}. \tag{21}$$

while for the RHP component

$$Q_+ = \begin{cases} m, & N > 2m \\ m - N, & N < 2m \\ 0, & N = 2m \end{cases}. \qquad (22)$$

The above expressions can be employed also in the case $r_0 \neq 0$. For example, for the array with $N = 3, m = 0$ the centered vortex topological charge in RHP component is $Q_+ = 0$ while in the LHP component we have $Q_- = -1$. Similarly to that for the array with $N = 5, m = 0$ we have $Q_+ = 0, Q_- = 2$, however the array with $N = 5, m = 1$ gives $Q_+ = 1, Q_- = -2$. The obtained expressions (21, 22) describe exactly the results of the computer simulations.

One cannot but say about the distortion of the centered vortex core in the beam array. The estimation presented in eqs (21) and (22) refers only to the phase increment while the path-tracing around the array axis but not to the vortex state. However, any deformations of the wave front in vicinity of the vortex core can result in the strong distortion of the vortex state up to its complete disappearance although the value $Q$ remains intact [34]. The vortex core deformation influences on the OAM. It means that we can treat the vortex deformation as appearing the additional optical vortex at the same position but with the opposite sign of the topological charge. Typical phase patterns shown in Fig.4 illustrate the core distortion in the centered vortex. For a relatively small number $N$ of the partial beams in the array a simple rule of changing the topological charge in the orthogonally polarized component by two units can experience the seeming violations (see, e.g., Fig.4 for $N = 3; 4; 5$). Moreover, the vortex core can be extremely deformed. However, as growing the number of the partial beams in the array, the distortions are smoothed over.

### II.3 Energy efficiency of the conversion process

As it is well known, the propagation of the circularly polarized plane wave perpendicular to the crystal optical axis is accompanied by the periodical transmission of the light energy between RHP and LHP components. Different versions of such a process are also observed in the paraxial beams propagating along the crystal optical axis [35], the maximum energy efficiency taking place for the Bessel-Gaussian beams. The base of the process is the spin-orbit coupling [36] that conditions fulfillment of the conservation law of the total angular momentum flux as a sum of the spin (SAM) and OAM fluxes [20]. In turn, it enables us to restrict ourselves the study of only one circularly polarized component (say, the $E_+$ component) [35]. In accordance with the approach presented in the papers [23, 33] and using the expressions (18) and (19) one rewrites the field (9) for the case $l = 0$ in the form:

$$E_+ = N \sum_{q=-\infty}^{\infty} \left(\frac{\Re - \bar{\alpha}}{\Re + \bar{\alpha}}\right)^{\frac{qN+m}{2}} e^{i(qN+m)\varphi} \left[I_{qN+m}(\xi_o)\Psi_o + I_{qN+m}(\xi_e)\Psi_e\right]. \quad (23)$$

Each term in the sum (23) are of the Bessel-Gaussian beam [25]. In the limit of $N \to \infty$, the terms with $q \neq 0$ give no contribution to eq. (23) and we can write down

$$E_+ \sim \left[(\Re + \bar{\alpha})/(\Re - \bar{\alpha})\right]^{m/2} \left[I_m(\xi_o)\Psi_o + I_m(\xi_e)\Psi_e\right]\exp(im\varphi). \quad (24)$$

In fact we deal with the sum of the ordinary and extraordinary Bessel-Gaussian beams. In the case of the self-matched beams (or spiral beams) $\Re = \bar{\alpha}$ and eq. (24) is transformed into the sum of the ordinary and extraordinary vortex-beams:

$$E_+ \sim \left[\Psi_o \sigma_o^{-m} + \Psi_e \sigma_e^{-m}\right] r^{|m|} \exp(im\varphi). \quad (25)$$

But it is the Bessel-Gaussian beams that have the energy efficiency of the conversion process nearly 100% [35]. Let us consider now the contribution of the phase-matched beam array contributes to such an exchange process. The value of the energy efficiency is given by the expression [35]:

$$\eta_{l,m} = \left\{\iint_{S_\infty} |E_-^{l,m}|^2 dS\right\} / \left(\iint_{S_\infty} |E_-^{l,m}|^2 dS + \iint_{S_\infty} |E_+^{l,m}|^2 dS\right), \quad (26)$$

where $S_\infty$ is the crystal's cross-section with the radius $R_0 \to \infty$. The curves shown in Fig.5 are of dependencies of the energy efficiency (26) as functions of the $LiNbO_3$ crystal length $z$ and the inclination angle $\alpha_o$. Small deviations of the crystal length $z$ and the array angle $\alpha_o$ entail fast oscillations of the efficiency $\eta$. However, for comparatively small lengths $z$ and angles $\alpha_o$, the energy efficiency of the vortex conversion can approach 100%. We have also revealed that the phase matched number $m$, the vortex topological charge $l$ and the number of local beams $N$ have almost no influence on the exchange while the waist $w_0$ of the local beams essentially change of the value $\eta$. The energy transport between the orthogonal components ceases quickly while growing the length $z$ or the angle $\alpha_o$ for relatively small values of the radii $w_0$ (see the lowest curves in Fig.5).

### III. Rotational spin-Hall effect

#### III.1 The model representation

When transmitting a paraxial beam with initial circular polarization (say, the RHP) and the $l$-charged centered vortex along the optical axis of a uniaxial crystal, the LHP component gains the additional two-charged optical vortex [27]. The transformation process is controlled by the conservation law of the angular momentum flux [37]. In the paraxial approximation the total angular momentum flux along the crystal optical axis can be presented as a sum of the OAM $L_z$ and the SAM $S_z(z=0)$ per photon:

$$L_z(z) + S_z(z) = L_z(z=0) + S_z(z=0). \tag{27}$$

As the polarized beam propagates along the crystal its spatial degree of polarization $P(z)$ tends to zero $P(z \to \infty) \to 0$. Its spin angular momentum vanishes $S_z(z \to \infty) \to 0$. In such a situation the total angular momentum in eq. (27) can be unchanged only if the OAM compensates the loss of the SAM at the expense of nucleating the centered doubly charged vortex in the LHP component. When tilting the beam axis relative to the crystal optical axis at a small angle the total angular momentum is as before conserved [20] while the doubly charged vortex leaves the beam [19, 21]. But the spatial depolarization of the tilted beam takes away a portion of the angular momentum too. The compensation of the loss is made up with the extrinsic OAM in the form of the lateral shift of the center of gravity in the LHP component relative to the inclination plane of the beam [20]. Generally speaking the lateral shift is also caused by the destructive interference between the ordinary and extraordinary beams (see eqs (5) and (6)) in both circularly polarized components. It varies within a broad range of values. But stating from some values of the inclination angle $\alpha_o$ and the crystal length $z = d$ the contribution of the beam overlapping to the lateral shift get vanished. The asymptotic value of the lateral shift then is equal to

$$\Delta x = -2s / (k_o \alpha_o), \qquad (28)$$

where $s = \pm 1$ stands for the handedness of the initial circular polarization. Thus the direction of the lateral shift depends neither on the type of the paraxial beam no on the topological charge of the vortex bearing the singular beam but is only defined by the handedness of the circular polarization of the initial beam.

All local beams in the LHP component of the array experience the same asymptotic displacement (28). For the case $r_0 = 0$, the angular displacement $\Delta \varphi$ of all local beams or, otherwise, the asymptotic rotational angle of the orthogonal polarized component of the beam array has a simple geometrical interpretation:

$$\Delta \varphi \approx \Delta x / r = -2s / (k_o \alpha_o \alpha_m z), \qquad (29)$$

where $\alpha_m = (\alpha_o + \alpha_e)/2$. Similar to the lateral shift (28), the asymptotic angular displacement of the beam array with a finite number of local beams (23) depends neither on the vortex topological charge $l$ nor on the phase-matching number $m$ but it is inversely proportional to the square of the inclination angle and tends to zero for a large crystal length.

Absolutely different situation occurs for the case of a great number of local beams, relatively small inclination angles and the crystal lengths. The beam overlapping here is the whole show. In this case we have used the following model. The beam cross section at the distance $z$ inside the crystal is divided into the $N$ sectors (see Fig.6). The center of each *n-th* local beam is positioned at the center of the *n-th* sector. Then we have calculated the center of gravity of the field intensity in the *n-th* sector both for the RHP and the LHP components:

$$\tan\varphi_n^{(\pm)} = \int\limits_{\varphi_n}^{\varphi_{n+1}}\int\limits_0^\infty |E_\pm|^2 r^2 \sin\varphi dr d\varphi \Big/ \int\limits_{\varphi_n}^{\varphi_{n+1}}\int\limits_0^\infty |E_\pm|^2 r^2 \cos\varphi dr d\varphi. \qquad (30)$$

The rotational angle is calculated as $\Delta\varphi = \varphi_n^{(+)} - \varphi_n^{(-)}$. Curves in Fig.7 illustrate the angular displacement $\Delta\varphi$ when changing the crystal length $z$. The array rotation has the oscillation features. At a small crystal length ($z \sim 1mm$) the angular displacement $\Delta\varphi$ can reach $-11^o \div 13^o$ but the oscillation amplitude descends quickly tending to zero. The oscillations in the array with a relatively large waist radius $w_0$ of the local beams fade out slower than those in the array with a relatively small $w_0$ (compare Fig.7A and 7B). The growth of number $N$ of local beams in the array shown in Fig.8 manifests itself in tending the angular displacement $\Delta\varphi$ to zero. The ordinary and extraordinary beam arrays in the RHP component are turned into the Bessel-Gaussian beams.

It is worth noting the asymmetry in the upper and the lowest envelopes of the curves $\Delta\varphi(z, \alpha_o = const)$ in Fig.7 and $\Delta\varphi(\alpha_o, z = const)$ in Fig.9A. In order to bring to light the role of the asymmetry in the rotation process let us remark that the destructive interference between neighbouring local beams and between the ordinary and extraordinary beams in each circularly polarized component results in the nucleation and annihilation of additional optical vortices. It stimulates displacements and distortions of the vortex cores in the eigen vortices [22] causing the transformation of the intrinsic and extrinsic OAM of the local beams [32] and the oscillation of the OAM in the beam array as a whole. In this case the upper and the lowest envelopes must be symmetrically positioned in the curves $\Delta\varphi(\alpha_o)$ and $\Delta\varphi(z)$. However there is one more effect that breaks the symmetry down. It is associated with the spatial depolarization of the beam appearing as the additional angular displacement of the beam. We have already derived the expression (29) for its asymptotic value. The decay of the SAM does not entail the oscillation of the angular displacement but decreases only the amplitude of the oscillations. Responsible for it is the polarization degree being the envelope of the SAM oscillations shown in Fig.9B (see also [20,35]. The polarization degree tends to zero more quickly than the overlapping between neighboring local beams or the ordinary and extraordinary beams in each local beam. At the same time, the vanishing of the SAM causes the angular displacement with the strongly defined handedness (see eq.(29)). As a result the amplitudes of the right and left rotations turn out to be different. The circlets in Fig.9A represent differences between the upper and the lowest envelopes of the curve $\Delta\varphi(\alpha_o)$. This values lie down very well on the theoretical asymptotic curve 2 of the angular displacement (29) of the beam array derived from the conservation law of the angular momentum. Such a good agreement takes place for not all values of the inclination angles $\alpha_o$ but only for those where the OAM $S_z \ll 1$. In particular for our case the comparison of Fig.9A and B gives the requirement: $\alpha_o > 2^o$. In the case of the LHP initial beam, the additional angular displacement changes its rotation direction for RHP component.

### III.2  The experiment

For the experimental study of the beam array evolution in a uniaxial crystal we have used the experimental set-up whose sketch shown in Fig.10. The shaping of the beam array was performed by the following way. The polarization filter consisting of the polarizer P and the quarter-wave retarder $\lambda/4$ transforms the initial Gaussian beam from He-Ne laser at the wavelength $\lambda = 0.6328\,\mu m$ into the circularly polarized beam. The spatial phase modulator converts the Gaussian beam into the vortex-beam with the topological charge $m = 1$. Then the vortex-beam passes through the conical lens – axicon with the apex-angle $\theta = 175^o$ being transformed into the Bessel-Gaussian-like beam. The corresponding intensity distribution of the beam cross-section after each unit of the experimental set-up is shown in Fig.10. Then the beam drops on the figured diaphragm with the N pinholes positioned at the vertices of the regular polygon shown in Fig.10. The lens system $L_1 L_2$ with two lenses with the focus distances $f_1$ and $f_2$ enables us to shape the required form of the beam with the defined inclination angle and the beam radius of the local beams at the input face of the $LiNbO_3$ crystal (the length $z = 2cm$). The output beam after the crystal is collimated by the lens system (not shown in Fig.10) and passes trough the polarization filter $\lambda/4, P$ that selects the RHP and LHP field components. Then the beam array is projected by the lens $L_3$ onto the input pupil of the CCD camera.

The shape of the array at the input face of the crystal is formed by changing the distance $d$ between the lenses $L_1$ and $L_2$, the initial shift $r_0$ of the local beams being zero $r_0 = 0$. Naturally, the change of the distance $d$ results in changing both the input angle of the local beams and their waist radii. To allow for such spatial transformations we can made use of the *ABCD* rule [38]. Indeed the symmetric beam array can be described in terms of the Bessel-Gaussian beams in eq. (23). Thus, to estimate theoretically the beam transformations in the two-lens experimental system it is sufficiently to apply the *ABCD* rule to the complex parameter $Z = z + i z_{o,e}$ in each Bessel-Gaussian beams in the sum. Although the approach did not permit us to describe aberrations of the local beams [39] we have achieved a good matching with our experiment. At the experiment we have recorded the distance $d$ between the lenses, the distance $\ell$ between the $L_2$ lens and the $LiNbO_3$ crystal, the inclination angle $\alpha_0$ of the local beams in the array and the size of the beam spot on the input face of the crystal, the inclination angles $\alpha_0$ before the crystal and $\alpha_o$ after the crystal being bound with each other by the simple relation $\alpha_0 \approx n_o \alpha_o$. The center of gravity of the sectors in the array (see Fig.6) at the experiment is estimated by the computer processing in accordance with the procedure described in the paper [20]. The experimental results shown in Fig.11 describe the evolution of the angular displacement $\Delta\varphi$ as a function of the inclination angle $\alpha_o$. The diameter of the pinhole at the figured diaphragm D is about $2\rho \approx 700\mu m$. In contrast to the quickly oscillating curves $\Delta\varphi(\alpha_o)$ presented in Fig.9A, the experimental curve in Fig.11 has only one maximum and one minimum. Such a strange behavior of the curve can be easily understood by recollecting that the variations of the distance $d$ entail both the change of the inclination angle $\alpha_o$ and the waist radius $w_0$ of the local beams at the crystal input. The range of variations of the waist radii $w_0$ at the crystal input is from $250 \mu m$ to $60 \mu m$. The partial oscillations $\Delta \varphi$ caused by each of these parameters are smoothed in the total complex wave process realized at the experiment. At relatively large values of the inclination angle $\alpha_0$ the angular displacement tends to zero. It's interesting that the additional angular displacement in our experiment can exceed one angular degree. The same process is reproduced by the theoretical curve having a good agreement with the experiment apart from a little displacement of the theoretical curve relative to the experimental one caused by the rough *ABCD* model that describes the experiment only approximately.

## IV Conclusions

We have theoretically and experimentally considered the propagation of the phase-matched array of singular beams through a uniaxial crystal indicating briefly its major features. We have revealed that the topological charge of the optical vortex nucleating at the array axis in the circularly polarized component orthogonal to that in the initial beam can reach the value $Q = m + 2 - N$ ($N$ is the number of local beams, $m$ stands for the phase-matching number). The optimal choice of the array parameters enables us to achieve nearly 100% energy efficiency of this conversion process.

We have also revealed the rotational spin Hall effect of the beam array that represents the additional angular rotation of the local beams in the array with the orthogonal circular polarization. In the general case there are two physical mechanisms that cause the destructive interference between the neighbouring local beams in the array and the interference between ordinary and extraordinary beam arrays in each polarized component. However these processes provide the symmetrical rotation with right and handedness of the array. But we have found asymmetry in the right and left rotations of the array that just gives the additional angular rotation. The value of the additional rotation can come up to some angular degrees. Such an angular asymmetry can be theoretically and experimentally separated from the total rotational process by the difference between the upper and lowest envelopes in the dependency of the angular rotation on the inclination angle of the local beams or on the crystal length. The angular asymmetry is caused by the spatial depolarization of the array and controlled by the conservation law of the total angular momentum flux along the crystal optical axis.

**FIGURES**

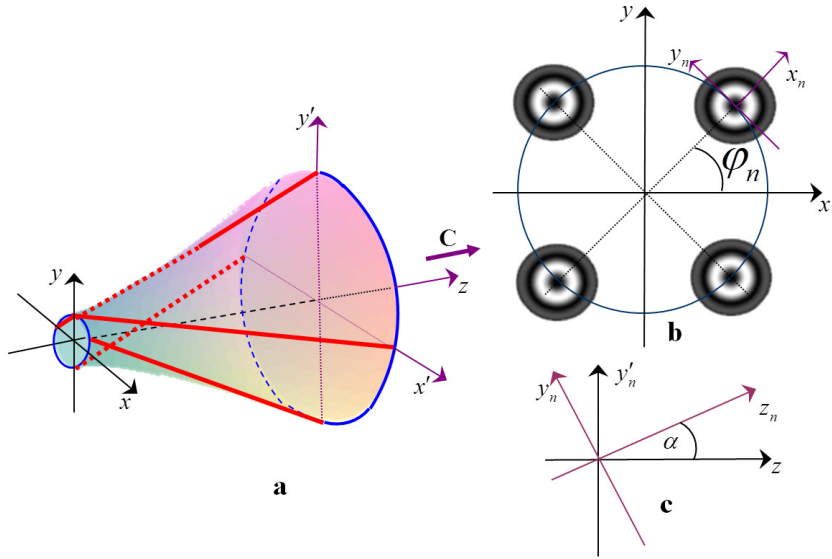

Fig.1 (Color online) Positions of the axes of partial beams (solid lines) on the surface of the hyperboloid of revolution; (a) positions of the partial beams at the initial plane $z=0$ (b); inclination of the beam axis (c); **C** is a unit vector of the crystal optical axis.

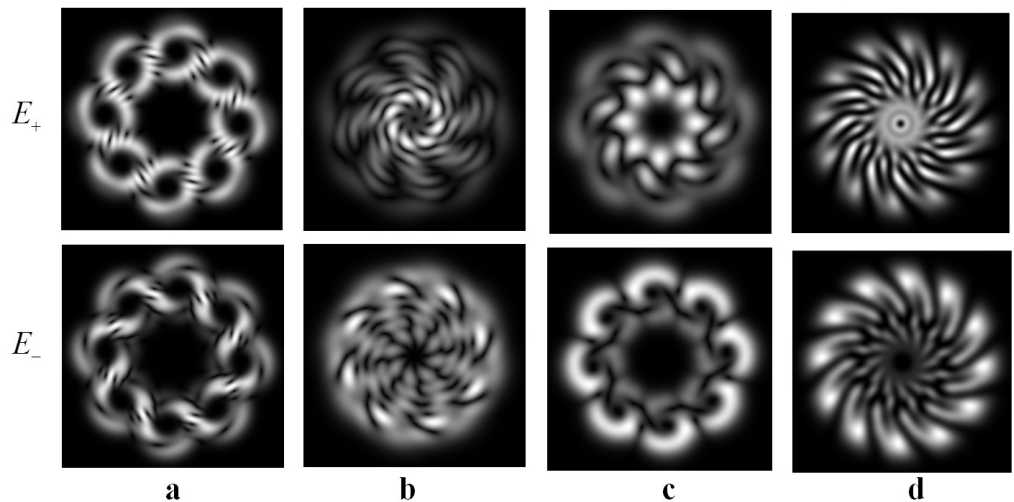

Fig.2 Intensity distributions in RHP and LHP components of different types of the phase-matched beam arrays in the $LiNbO_3$ crystal with $n_o = 2.2$, $n_e = 2.3$, $z = 2\,cm$: (a,b) $\{8,1,6\}$, $r_0 = 30\,\mu m, w_0 = 30\,\mu m$, (a) $\alpha_o = 1^o$, (b) $\alpha_o = 0.5^o$ ;(c) $\{8,1,6\}$ $r_0 = w_0 = 30\,\mu m, \alpha_o = 0.57^o$ (d) $\{12,1,4\}$, $w_0 = 50\,\mu m$, $r_0 = 50\,\mu m, \alpha_o = 0.54^o$

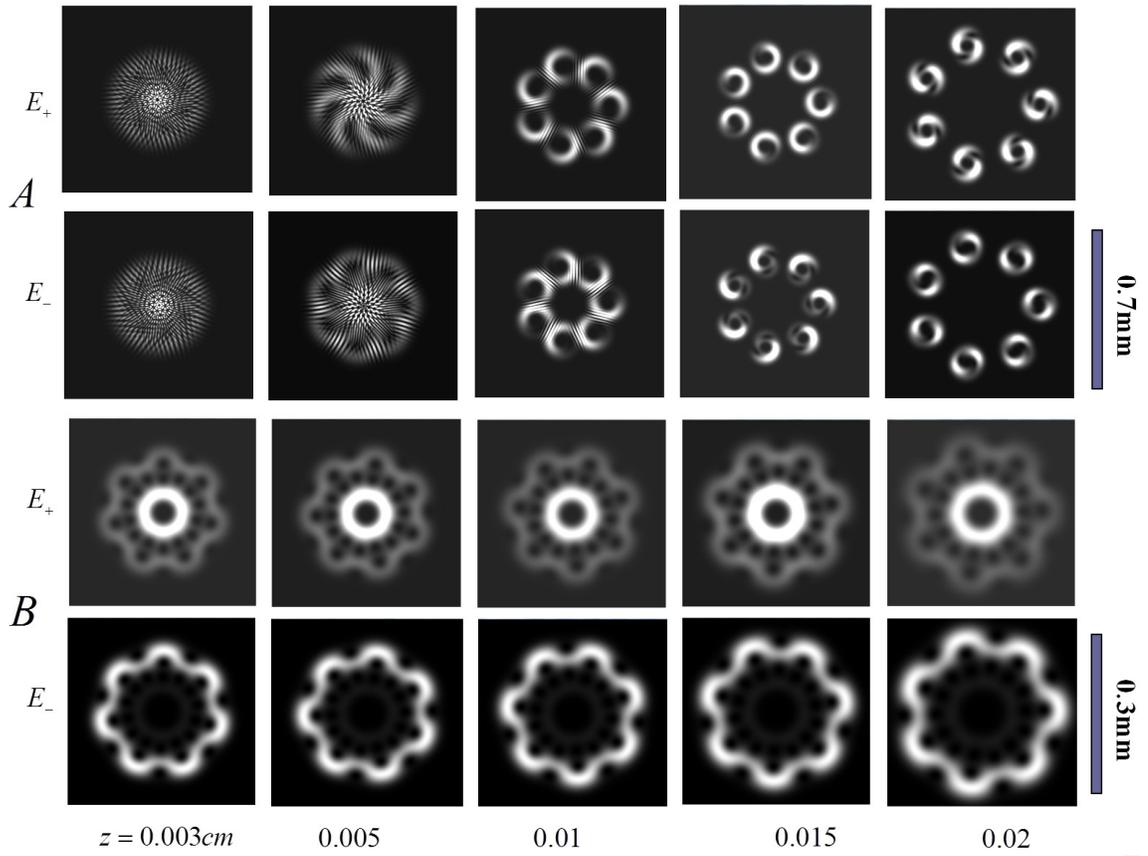

Fig.3 Propagation of the singular beam array along the uniaxial crystal with $N=7$, $m=3$, $l=4$, $w_0=50\mu m$, $n_o=2.2$, $n_e=2.3$: (A) $r_0=0$, $\alpha_o=1.2^o$; (B) spiral beam $\alpha_o=0.244^o$, $r_0=50\mu m$

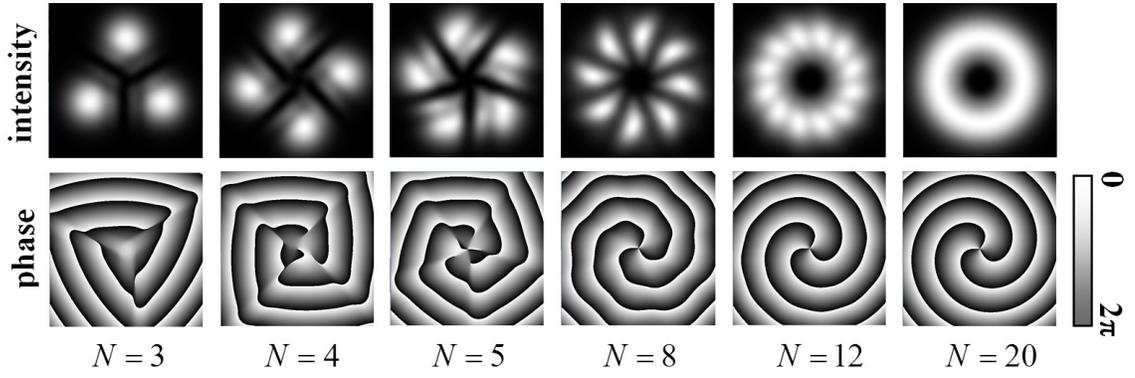

Fig.4 Deformation of the centered optical vortex in the LHP component $E_-$ of the phase-matched beam array with different number $N$ of the partial beams: $m=1$, $l=0$, $r_0=0$, $\alpha_o = 0.3^o$, $z = 1 cm$

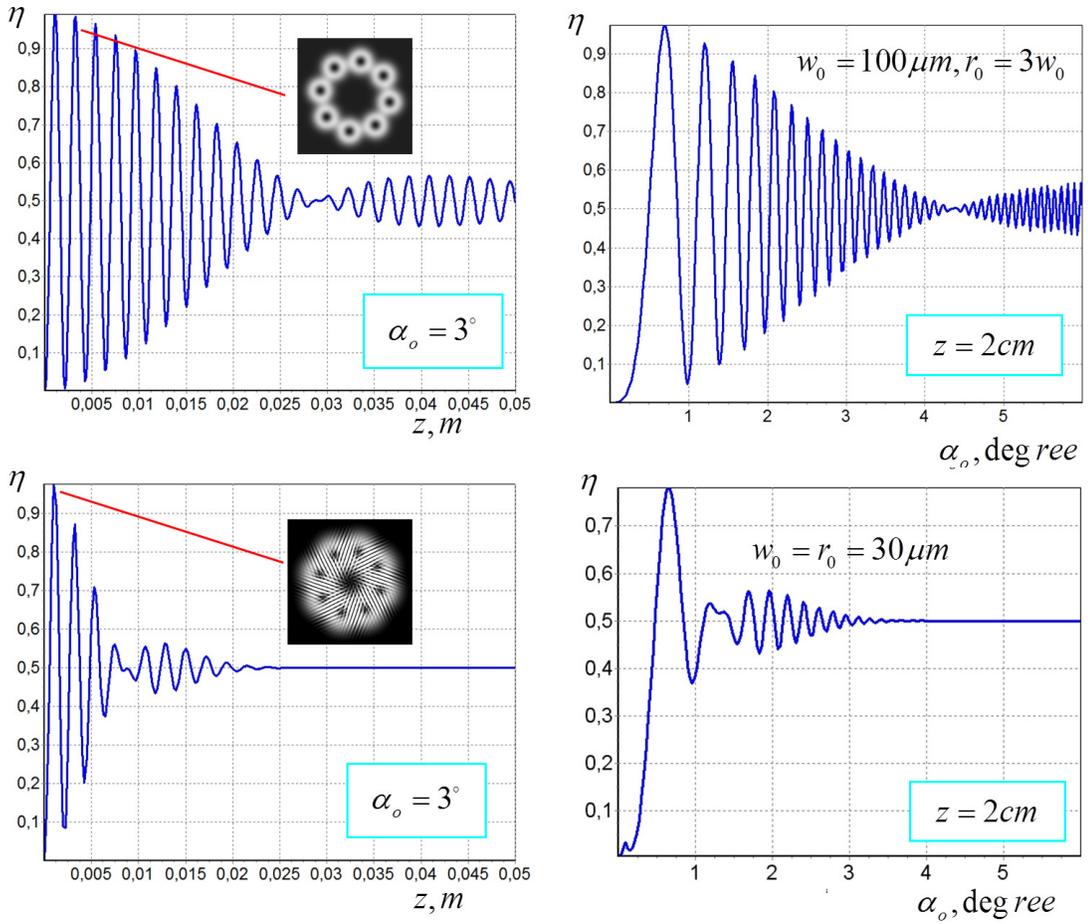

Fig. 5 (Color online) Energy efficiency $\eta(z)$ and $\eta(\alpha_o)$ for the beam array $\{8,1,1\}$ and corresponding intensity distributions

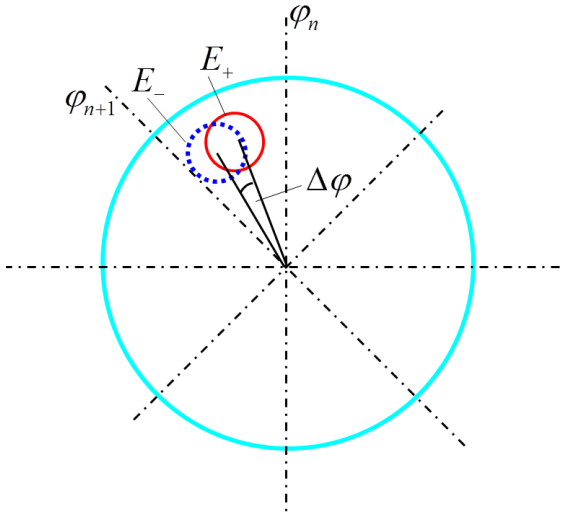

Fig.6 (Color online) The rotation of the beam array

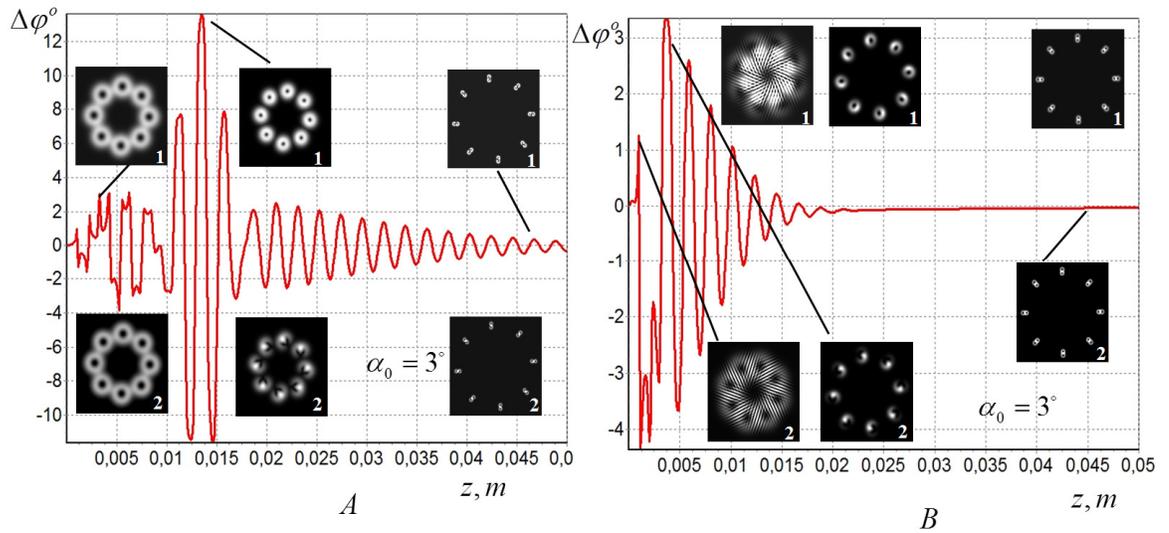

Fig.7 (Color online) Angular rotation $\Delta\varphi(z)$ of the array in the state $\{8,1,1\}$ with (A) $w_0 = 100\mu m, r_0 = 300\mu m$ and (B) $w_0 = r_0 = 30\mu m$. The designations (1) and (2) corresponds the field components $E_+$ and $E_-$, respectively, for the same values of the crystal length $z$

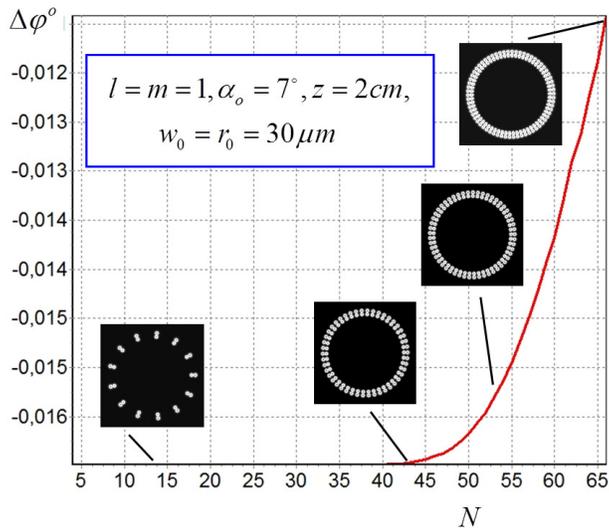

Fig.8 (Color online) The dependency of the angular displacement $\Delta\varphi$ on the number of local beams $N$ in the array.

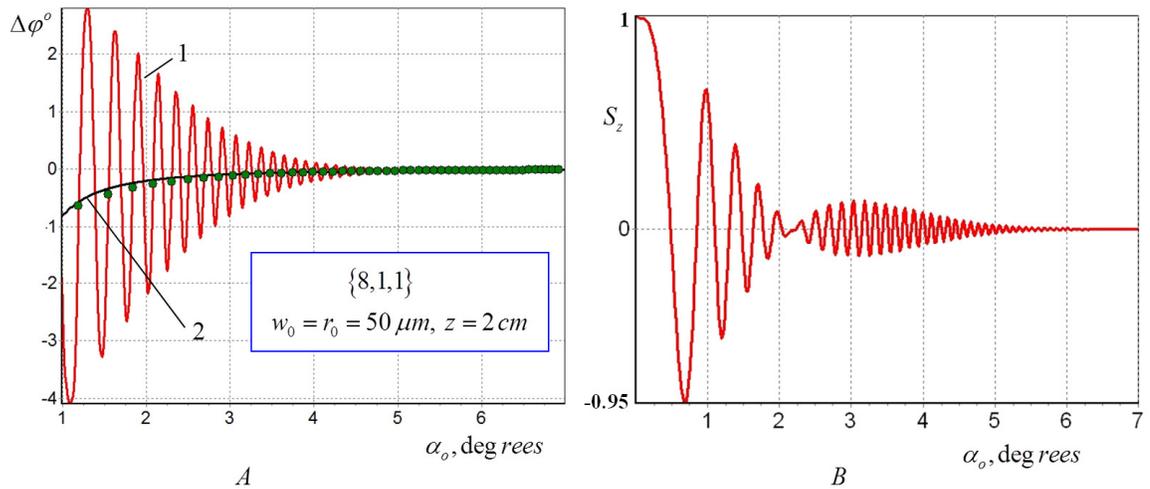

Fig.9 (Color online) (A) Angular displacement $\Delta\varphi$ as a Fig.9 (A) Angular displacement $\Delta\varphi$ as a function of the inclination angle $\alpha_o$ (1) and (2) the asymptotic curve (23); *the circlets* are of the difference between the upper and the lowest envelopes of the curve 1 (in all 800 circlets); (B) evolution of the SAM $S_z(\alpha_o)$

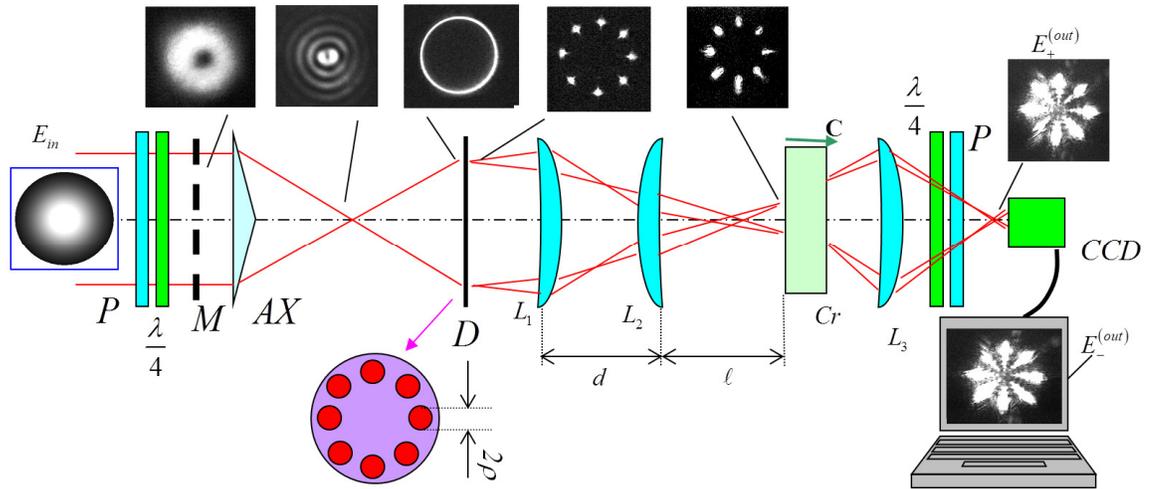

Fig.10 (Color online) Sketch of the experimental set-up: P – polarizer, λ/4 – quarter-wave retarder, M – spatial phase modulator, AX – axicon, D – figured diaphragm, $L_{1,2,3}$ – lenses, Cr - $LiNbO_3$ crystal

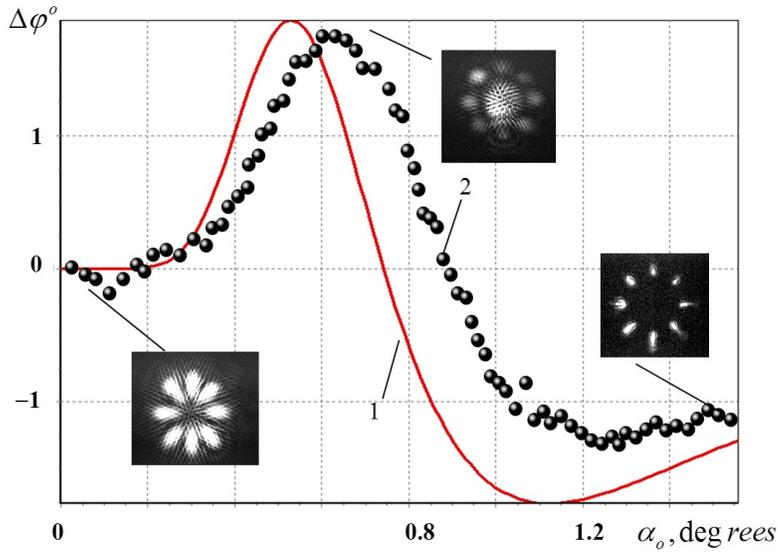

Fig. 11 (Color online) Theoretical (1) and experimental (2) curves $\Delta\varphi(\alpha_o)$ and corresponding intensity distributions in the LHP array component.